\documentclass[prb]{revtex4}

\usepackage{bm}

\begin{document}
\title{D'yakonov-Perel' spin relaxation under electron-electron
collisions in QWs}

\author{M.M.~Glazov and E.L.~Ivchenko}

\affiliation{A.F.~Ioffe Physico-Technical Institute, Russian
Academy of Sciences, 194021 St.~Petersburg, Russia}
\begin{abstract}

The D'yakonov-Perel' mechanism of spin relaxation is connected
with the spin splitting of the electron dispersion curve in
crystals lacking a center of symmetry. In a two-dimensional
noncentrosymmetric system, e.g. quantum well or heterojunction,
the spin splitting is a linear function of ${\bm k}$, at least for
small values of ${\bm k}$. We demonstrate that the spin relaxation
time $\tau_s$ due to the spin splitting is controlled not only by
momentum relaxation processes as widely accepted but also by
electron-electron collisions which make no effect on the electron
mobility. In order to calculate the time $\tau_s$ taking into
account the electron-electron scattering we have solved the
two-dimensional kinetic equation for the electron spin density
matrix. We show how the theory can be extended to allow for
degenerate distribution of the spin-polarized two-dimensional
electron gas.
\end{abstract}
\maketitle
\section{Introduction}
In recent years the electron spin relaxation processes have
received much attention in connection with various spintronics
applications. For free electrons in quantum well structures the
following four mechanisms of spin decoherence are most important
(see Refs.~\onlinecite{book,agw_rev} and references therein): the
Elliot-Yafet, D'yakonov-Perel',  Bir-Aronov-Pikus mechanisms and
electron spin-flip scattering by paramagnetic centers. This paper
is devoted to the D'yakonov-Perel' spin relaxation (DPSR) in which
case the spin relaxation time, $\tau_s$, is related to the spin
splitting and given by Ref.~\onlinecite{dp}
\begin{equation} \label{taus}
\tau_s^{-1} \propto \langle {\bm \Omega}^2_{\bm k} \rangle \tau
\:.
\end{equation}
Here ${\bm \Omega}_{\bm k}$ is the effective Larmor frequency
describing the splitting of the electron spin dispersion branches,
the angle brackets mean averaging over the electron energy
distribution and $\tau$ is a microscopic electron relaxation time.
In a two-dimensional (2D) system laking a center of a symmetry,
e.g. quantum well or heterojunction, the frequency ${\bm
\Omega}_{\bm k}$ is linear in ${\bm k}$. In this case the time
$\tau$ was usually identified with the momentum relaxation time
that determines the electron mobility
Refs.~\onlinecite{dk,ivchenko2,ohno1,ohno2,harley}. In
Ref.~\onlinecite{gi} we have shown that the inverse time
$\tau^{-1}$ is determined not only by the momentum scattering rate
but contains also an independent contribution from
electron-electron collisions which make no effect on the electron
mobility. Really, electron-electron collisions change the
direction of ${\bm k}$ and ${\bm \Omega}_{\bm k}$ and, therefore,
they control the Dyakonov-Perel' spin relaxation exactly in the
same way as any other scattering processes do.

The paper is organized as follows. Section 2 contains the
discussion of mechanisms of the spin splitting of the electron
subbands in quantum wells and electron spin decoherence due to
spin splitting, Sec. 3 describes the D'yakonov-Perel' spin
relaxation mechanism in terms of the kinetic theory, in Sec. 4 we
analyze the electron-electron collision integral for the
spin-polarized electron gas, and Sec. 5 describes the solution of
the kinetic equation with the account of electron-electron
collisions.
\section{Spin Splitting of Electron Subbands}
In the parabolic approximation the effective electron Hamiltonian
in the zinc-blende-based $(001)$-grown quantum well (QW) can be
written as
\begin{equation}\label{hamiltonian}
{{\cal{H}}} = \frac{\hbar^2 k^2}{2 m} + \beta_1 (\sigma_y k_y -
\sigma_x k_x) + \beta_2 (\sigma_x k_y - \sigma_y k_x)\:,
\end{equation}
where $x\|[100]$, $y\|[010]$, $\beta_{1,2}$ are constants, $m$ is
the in-plane effective mass, $\sigma_l$ are the Pauli matrices and
$k_n$ are the components of the 2D electron wavevector, $\bm k$.
The terms with higher powers in $\bm k$ are not considered here.
In the symmetrical QWs (the $D_{2d}$ point group) the linear-${\bm
k}$ spin-dependent term proportional to $\beta_1$ is allowed only,
it is called the bulk-inversion asymmetry (BIA) term. In
asymmetrical QW structures characterized by the $C_{2v}$
point-group symmetry, there exists another spin-dependent
contribution proportional to $\beta_2$ and called the
structure-inversion asymmetry (SIA) term or the Rashba term
Refs.~\onlinecite{rashba1,rashba2} (see also Ref.~\onlinecite{agw}
and references therein). The structure asymmetry can be related
with non-equivalent normal and inverted interfaces, external or
built-in electric fields, compositionally stepped QWs etc.

It is convenient to introduce the Cartesian coordinates
$x'\|[1\bar{1}0]$, $y'\|[110]$, $z\|[001]$ which allow to write a
sum of the BIA and SIA terms in the form
\begin{equation}\label{hc1}
{{\cal H}}_{c1}({\bm k}) = \left( \beta_- \sigma_{x'} k_{y'} +
\beta_+ \sigma_{y'} k_{x'} \right)/2\:,
\end{equation}
where $\beta_{\pm}=2(\beta_2\pm\beta_1)$. The effective Larmor
frequency $\bm\Omega_{\bm k}$ is defined by ${{\cal H}}_{c1} =
(\hbar/2)\bm\Omega_{\bm k}\cdot \bm\sigma$, its components are
\begin{equation}\label{larmor}
\Omega_{ {\bm k}, x'} = \beta_-k_{y'} / \hbar\:,\quad \Omega_{{\bm
k}, y'} = \beta_+ k_{x'} /\hbar\:,\quad \Omega_{{\bm k}, z} = 0.
\end{equation}
The spin splitting at ${\bm k} = (k_{x'}, k_{y'})$ is
$\hbar\Omega_{\bm k} = \sqrt{\beta_+^2 k_{x'}^2+\beta_-^2
k_{y'}^2}$.  If only one of the linear-$\bm k$ terms, BIA or SIA,
is nonzero then $\beta_-^2 = \beta_+^2 \equiv \beta^2$ and the
splitting $\hbar \Omega_{\bm k} = \beta k$ is angular independent.

Consider an electron gas occupying the lowest conduction subband
$e1$ and assume that, at the moment $t=0$, the electrons are
spin-polarized in the same direction along, say, the growth axis
$z$. Owing to the spin-splitting of the electron subband the
electron spin in the state with the wavevector ${\bm k}$ precesses
around the axis ${\bm \Omega}_{\bm k}$ which lies in the interface
plane, see Eq.~(\ref{larmor}). In case of the large splitting,
$|{\bm \Omega}_{\bm k}| \gg 1 / \tau$, the spin of electron in the
state ${\bm k}$ will oscillate with time as $s_z(t) = s_z(0)
\cos{\Omega_{\bm k} t}$. In the case of ${\Omega}_{\bm k}$ being
isotropic in the QW plane, the spin polarized electrons which
occupy the circle of the fixed radius in the ${\bm k}$ space show
the similar oscillatory behavior for their $s_z$ component.
However, if $\beta_-^2 \ne \beta_+^2$ and/or the electrons occupy
states with different values of $|{\bm k}|$, the scatter in ${\bm
\Omega}_{\bm k}$ results in a fast non-exponential decay of $s_z$.

D'yakonov and Perel' Ref.~\onlinecite{dp} were the first to show
that the processes of electron-momentum scattering result in the
slowing off the spin decoherence caused by the spin splitting. In
the collision-dominated (``motional-narrowing") limit $|{\bm
\Omega}_{\bm k}| \ll 1/ \tau$, this results in an exponential
decay of the average spin: $\bar s_z(t) = \bar s_z(0) \exp{(- t /
\tau_s )}$, where $\tau_s$ is given by Eq. (\ref{taus}). The
dimensionless coefficient in Eq. (\ref{taus}) can be obtained from
the solution of kinetic equation Refs.~\onlinecite{dp,dk}. This
approach is valid as long as the energy relaxation time
$\tau_{\varepsilon}$ is short as compared to $\tau_s$.
\section{DPSR in Terms of the Kinetic Theory}
In the frame of kinetic theory, the electron distribution in the
wavevector and spin spaces is described by a $2 \times 2$
spin-density matrix
\begin{equation}
\hat{\rho}_{\bm k} = f_{\bm k} + {\bm s}_{\bm k} \cdot {\bm
\sigma}.
\end{equation}
Here $f_{\bm k} = \mbox{Tr}{ (\hat{\rho}_{\bm k} /2)}$ is the
average occupation of the two spin states with the wavevector
${\bm k}$, or distribution function of electrons in the $\bm
k$-space, and the average spin in the ${\bm k}$ state is ${\bm
s}_{\bm k}= \mbox{Tr}{[\hat{\rho}_{\bm k} ({\bm \sigma}/2)]}$. If
we neglect the spin splitting then, for arbitrary degeneracy of an
electron gas with non-equilibrium spin-state occupation but
equilibrium energy distribution within each spin branch, the
electron spin-density matrix can be presented as
\begin{equation} \label{rho0deg}
\hat{\rho}^0_{\bm k} = \left\{ \exp{ \left[ \frac{ E_{\bm k} -
\bar{\mu} - \tilde{\mu}\: ({\bm \sigma} {\bf o}_{\bm s}) }{k_B T}
\right] } + 1 \right\}^{-1}\:,
\end{equation}
where $E_{\bm k} = \hbar^2 k^2 /2m$, $k_B$ is the Boltzmann
constant, $T$ is the temperature, ${\bf o}_{\bm s}$ is the unit
vector in the spin polarization direction, $\mu_{\pm} = \bar{\mu}
\pm \tilde{\mu}$ are the effective Fermi energies for electrons
with the spin component 1/2 or $- 1/2$ along ${\bf o}_{\bm s}$ so
that the energy distribution functions of electrons with the spin
$\pm 1/2 $ are given respectively by
\[
f_{k,\pm} = \left[ \exp{ \left( \frac{ E_{\bm k} - \mu_{\pm} }{k_B
T} \right) } + 1 \right]^{-1} \:.
\]
Note that Eq.~(\ref{rho0deg}) can be rewritten in the equivalent
form Ref.~\onlinecite{dp}
\[
\hat{\rho}^0_{\bm k} \equiv f^0_{k} + {\bm s}^0_{k} \cdot {\bm
\sigma} = \frac12 \left[ f_{k,+} + f_{k,-} + (f_{k,+} - f_{k,-})\:
({\bm \sigma} {\bf o}_{\bm s}) \right]\:.
\]
The densities $n_{\pm}$ of 2D electrons with a particular spin can
be related with the effective Fermi energies by
\begin{equation} \label{n+-}
n_{\pm} = \frac{m}{2 \pi \hbar^2}\: k_B T \ln{(1 +
e^{\mu_{\pm}/k_B T })}\:.
\end{equation}

If the spin splitting is non-zero but small compared to $\hbar /
\tau$, the distribution function $\mbox{Tr}{[\rho_{\bm k}/2]} = f_{\bm
k}^0$ does not change, whereas the spin vector obtains a
correction $\delta {\bm s}_{\bm k} = {\bm s}_{\bm k} - {\bm
s}^0_{k}$ proportional to the spin splitting. Therefore, the
spin-density matrix may be presented as
\begin{equation} \label{rho}
\hat{\rho}_{\bm k} = \hat{\rho}^0_{\bm k} + \delta {\bm s}_{\bm k}
\cdot {\bm \sigma}\:.
\end{equation}

The quantum kinetic equation for the spin-density matrix taking
into account the electron-electron collisions has the form
\begin{equation}\label{rho_eq}
\frac{\partial  \hat \rho_{\bm k}}{\partial t} + \frac{i}{\hbar} [
\mathcal H_{c1}(\bm{k}), \hat \rho_{\bm k}] + \hat{Q}_{\bm
k}\{\hat \rho \} = 0\:,
\end{equation}
where $[P,R] = PR - RP$, $\mathcal H_{c1}(\bm{k})$ is the
linear-${\bm k}$ Hamiltonian (\ref{hc1}), and the third term in
the left-hand side is the collision integral or the scattering
rate, in this equation it is a 2$\times$2 matrix. It follows from
Eq.~(\ref{rho_eq}) that the pseudovector kinetic equation for ${\bm
s}_{\bm k}$ can be written as
\begin{equation} \label{kinetic}
\frac{d {\bm s}_{\bm k}}{dt} + {\bm \Omega}_{\bm k} \times {\bm
s}_{\bm k} + {\bm Q}_{\bm k} \{ {\bm s} \} = 0\:,
\end{equation}
where ${\bm Q}_{\bm k} \{ {\bm s}, f \}$ = (1/2)Tr$({\bm
\sigma}\hat{Q}_{\bm k}\{ \rho \})$. In the following we ignore
spin flips under scattering. Then, say, for the elastic scattering
one has
\[
{\bm Q}_{\bm k} \{ {\bm F} \} = \sum_{{\bm k}'} W_{{\bm k}' {\bm
k}} ({\bm F}_{\bm k} - {\bm F}_{{\bm k}'})\:,
\]
where $W_{{\bm k}' {\bm k}}$ is the probability rate for the
electron transition from the state ${\bm k}$ to ${\bm k}'$. The
collision integral for electron-electron scattering is considered
in the next section. Here it suffices to note that, for the
distribution $\hat{\rho}^0_{\bm k}$, the collision integral
vanishes identically. This integral also vanishes after the
summation over ${\bm k}$ which allows, in particular, to derive
from Eq.~(\ref{kinetic}) the following equation of balance for the
total average spin ${\bm S}_0 = {\bf o}_{\bm s} (n_+ - n_-)$:
\begin{equation} \label{balances0}
\frac{d {\bm S}_0}{dt} + \sum_{\bm k} {\bm \Omega}_{\bm k} \times
\delta {\bm s}_{\bm k} = 0\:.
\end{equation}
The angular dependence of the non-equilibrium correction $\delta
{\bm s}_{\bm k}$ is the linear combination of $\cos{\Phi_{\bm k}}=
k_{x'}/k$ and $\sin{\Phi_{\bm k}} = k_{y'}/k$, where $\Phi_{\bm
k}$ is the angle between ${\bm k}$ and the axis $x'$. Retaining in
the kinetic equation (\ref{kinetic}) terms proportional to the
first angular harmonics we obtain the equation for $\delta {\bm
s}_{\bm k}$ with the inhomogeneous term linear in ${\bm S}_0$.
Then one can substitute the solution in the second term of
Eq.~(\ref{balances0}). The final result is that the tensor of
inverse spin relaxation times, $1/ \tau^s_{\alpha \beta}$, is
diagonal in the coordinate system $x', y', z$ and given by
Refs.~\onlinecite{dk,ag}
\begin{equation} \label{srtimes}
\frac{1}{\tau^s_{x'x'}} = \frac12 \left( \frac{\beta_+}{\hbar}
\right)^2 \langle k^2 \tau_p \rangle\:,\: \qquad
\frac{1}{\tau^s_{y'y'}} = \frac12 \left( \frac{\beta_-}{\hbar}
\right)^2 \langle k^2 \tau_p \rangle\:,\qquad
\frac{1}{\tau^s_{zz}} = \frac{1}{\tau^s_{x'x'}} +
\frac{1}{\tau^s_{y'y'}} \:,
\end{equation}
where $\tau_p$ is the momentum relaxation time. If among the two
contributions, BIA and SIA, to the spin splitting one is dominant
and $|\beta_+| = |\beta_-|$, the spin relaxation times are
interconnected by Ref.~\onlinecite{dk}\[ \tau^s_{x'x'} =
\tau^s_{y'y'} = 2 \tau^s_{zz}\:.
\]
Interplay between the BIA and SIA contributions can lead to a
giant spin relaxation anisotropy Ref.~\onlinecite{ag}. In
particular, if these contributions coincide, $\beta_1 = \beta_2$,
so that $\beta_- = 0$ one has $\tau^s_{x'x'} = \tau^s_{zz}$ and
$\tau^s_{y'y'} = \infty$. In the case $\beta_1 = - \beta_2$ the
coefficient $\beta_+ = 0$, the time $\tau^s_{x'x'}$ is infinite
and $\tau^s_{y'y'}$ coincides with $\tau^s_{zz}$.
\section{Electron-Electron Collisions in QWs}
Here we will write the electron-electron collision term
$\hat{Q}_{\bm k}\{ \hat \rho \}$ in the general case of arbitrary
spin-density matrix $\hat \rho_{\bm k}$ (in particular, arbitrary
degeneracy and arbitrary distribution of spin in the ${\bm
k}$-space). For this purpose we remind that the matrix element of
the Coulomb scattering ${\bm k}, s_{{\bm k}}$ + ${\bm k}', s_{{\bm
k}'}$ $\rightarrow$ ${\bm p}, s_{{\bm p}}$ + ${\bm p}', s_{{\bm
p}'}$ is given by
\begin{equation} \label{matcoul}
M({\bm p}, s_{\bm p}; {\bm p}', s_{{\bm p}'}| {\bm k}, s_{\bm k};
{\bm k}', s_{{\bm k}'}) = V_{\bm k - \bm p}\: \delta_{s_{\bm p},
s_{\bm k}} \delta_{s_{{\bm p}'}, s_{{\bm k}'}}- V_{\bm k - {\bm
p}'}\: \delta_{s_{\bm p}, s_{{\bm k}'}} \delta_{s_{{\bm p}'},
s_{\bm k}}\:,
\end{equation}
where $s_{\bm k}, s_{{\bm k}'}...= \pm 1/2$, $V_{\bm q}$ is a
Fourier transform of the 2D Coulomb potential of the
electron-electron interaction
\begin{equation} \label{2d}
V_{\bm q} = \frac{2\pi e^2}{\ae q{\rm \Sigma}}\:,
\end{equation}
$e$ is the elementary charge, $\ae$ is the dielectric constant,
and ${\rm \Sigma}$ is the sample area in the interface plane; in
the following we set ${\rm \Sigma} = 1$. Equation~(\ref{matcoul})
takes into account both the direct and exchange Coulomb
interaction.

In order to present $\hat{Q}_{\bm k}\{\hat \rho \}$ in a compact
form we introduce the 2$\times$2 unit matrix $I^{(1)}$ and Pauli
matrices $\sigma^{(1)}_{\alpha}$ ($\alpha = x,y,z$) for the spin
coordinates $s_{\bm p}, s_{\bm k}$ and a similar set of four
matrices, $I^{(2)}$ and $\sigma^{(2)}_{\alpha}$, for the spin
coordinates $s_{{\bm p}'}, s_{{\bm k}'}$. One can check that
Eq.~(\ref{matcoul}) allows the following matrix representation
\begin{equation} \label{mab}
\hat{M} = A \:I^{(1)}I^{(2)} + B\: \bm{\sigma}^{(1)} \cdot
\bm{\sigma}^{(2)}\:,
\end{equation}
\begin{equation} \label{ab}
A=V_{\bm k - \bm p} - \frac12\:V_{\bm k - \bm p'}\:,\: B = -
\frac12\: V_{\bm k - \bm p'}\:.
\end{equation}
Now the collision term for the electron spin-density matrix can be
presented as
\begin{equation} \label{qgeneral}
\hat{Q}_{\bm k}\{ \rho \} = \frac{\pi}{2 \hbar} \sum_{{\bm k}'
{\bm p} {\bm p}'} \delta_{{\bm k} + {\bm k}',\: {\bm p} + {\bm
p}'} \: \delta(E_k + E_{k'} - E_p - E_{p'}) \:\mbox{Tr}_2
G(\bm{p},\bm{p}'; \bm{k}, \bm{k}')\:,
\end{equation}
\begin{eqnarray} \label{green}
&&G(\bm{p},\bm{p}'; \bm{k}, \bm{k}') =\\&=& \hat{M} ( I^{(1)} -
\hat \rho^{(1)}_{\bm{p}} ) ( I^{(2)} - \hat \rho^{(2)}_{\bm{p}'} )
\hat{M} \hat \rho^{(1)}_{\bm{k}} \hat \rho^{(2)}_{\bm{k}'}  + \hat
\rho^{(1)}_{\bm{k}} \hat \rho^{(2)}_{\bm{k}'} \hat{M} ( I^{(1)} -
\hat \rho^{(1)}_{\bm{p}} ) ( I^{(2)} - \hat \rho^{(2)}_{\bm{p}'} )
\hat{M} -\nonumber \\ &-& \hat{M} \hat \rho^{(1)}_{\bm{p}} \hat
\rho^{(2)}_{\bm{p}'} \hat{M} ( I^{(1)} - \hat \rho^{(1)}_{\bm{k}}
) ( I^{(2)} - \hat \rho^{(2)}_{\bm{k'}} ) - ( I^{(1)} - \hat
\rho^{(1)}_{\bm{k}} ) ( I^{(2)} - \hat \rho^{(2)}_{\bm{k'}} )
\hat{M} \hat \rho^{(1)}_{\bm{p}} \hat \rho^{(2)}_{\bm{p}'} \hat{M}
\:. \nonumber
\end{eqnarray}
Here the spin-density matrices $\hat \rho^{(1)}(\bm{k}) = I^{(1)}
f_{\bm k} + {\bm \sigma}^{(1)} \cdot {\bm s}_{\bm k}$, $\hat
\rho^{(2)}(\bm{k}') = $ $I^{(2)} f_{{\bm k}'} + {\bm \sigma}^{(2)}
\cdot {\bm s}_{{\bm k}'} $ etc., the symbol Tr$_2$ means the trace
over the spin variable 2. After the trace is found the index 1 in
Tr$_2$$G(\bm{p},\bm{p}'; \bm{k}, \bm{k}')$ can be omitted. In
order to derive Eqs.~(\ref{qgeneral}), (\ref{green}) we used the
standard diagram technique.

Instead of equation (\ref{rho_eq}) for the spin-density matrix one
can use a scalar equation for the distribution function $f_{\bm k}
$ in the form
\begin{equation}
\frac{d{f}_{\bm k}}{d t} + {Q}_{{\bm k}} \{f, \bm{s} \} = 0
\end{equation}
and a equation for the spin-distribution vectorfunction as
\begin{equation}
\frac{d\bm{s}_{\bm k}}{d t} + \bm{\Omega}_{\bm k} \times
\bm{s}_{\bm k} + \bm{Q}_{{\bm k}} \{ \bm{s}, f \} = 0\:,
\end{equation}
where
\begin{equation}\label{qf-general}
Q_{\bf k} \{f , \bm s \} = \frac12 \mbox{Tr}_1 [ \hat{Q}_{\bm k}\{
\hat \rho \} ]
\end{equation}
\[
= \frac{\pi}{4\hbar} \sum_{{\bm k}' {\bm p} {\bm p}'}\delta_{{\bm
k} + {\bm k}',\: {\bm p} + {\bm p}'}\: \delta(E_k + E_{k'} - E_p -
E_{p'}) \: \mbox{Tr}_1 \mbox{Tr}_2 [ G(\bm{p},\bm{p}'; \bm{k}, \bm{k}') ] \:.
\]
\begin{equation}\label{q_common}
\bm{Q}_{\bf k} \{\bm s, f\} = \frac12 \mbox{Tr}_1 [
\bm{\sigma}^{(1)} \hat{Q}_{\bm k}\{ \hat \rho \} ]
\end{equation}
\[
= \frac{\pi}{4\hbar} \sum_{{\bm k}' {\bm p} {\bm p}'}\delta_{{\bm
k} + {\bm k}',\: {\bm p} + {\bm p}'}\: \delta(E_k + E_{k'} - E_p -
E_{p'}) \: \mbox{Tr}_1 \mbox{Tr}_2 [ \bm{\sigma}^{(1)} G(\bm{p},\bm{p}';
\bm{k}, \bm{k}') ] \:.
\]
For the analysis of the general equations (\ref{rho_eq}),
(\ref{qgeneral}), (\ref{qf-general}), (\ref{q_common}) we consider
below few particular cases.

{\it Spin-unpolarized electrons}. In this case $\bm{s}_{\bm k}
\equiv 0$ and the spin-density matrix reduces to a product of the
unit 2$\times$2 matrix and the distribution function $f_{\bm k}$.
Taking into account that
\[
\hat{M}^2 = (A^2 + 3 B^2) \:I^{(1)}I^{(2)} + 2 B (A - B)\:
\bm{\sigma}^{(1)} \cdot \bm{\sigma}^{(2)}
\]
we come to the conventional collision term
\begin{equation} \label{qf}
Q_{\bf k} \{f \} = \frac{2 \pi}{\hbar} \sum_{{\bm k}' {\bm p} {\bm
p}'}\delta_{{\bm k} + {\bm k}',\: {\bm p} + {\bm p}'}\: \delta(E_k
+ E_{k'} - E_p - E_{p'}) \:(A^2 + 3 B^2)
\end{equation}
\[
\times [f_{\bm k}f_{{\bm k}'}(1-f_{\bm p}) (1-f_{{\bm p}' }) -
f_{\bm p}f_{{\bm p}' }(1-f_{\bm k}) (1-f_{{\bm k}' })]\:.
\]
Note that according to Eq.~(\ref{ab}) one has
\[
A^2 + 3 B^2 = V_{\bm k - \bm p}^2 +  V_{\bm k - {\bm p}'}^2 -
V_{\bm k - \bm p}V_{\bm k - {\bm p}'}
\]
which is one-forth of the function $R$ introduced in Eq.~(2.4b) in
Ref.~\onlinecite{lyo} and the above collision term agrees with the
equation (2.4a) in the cited paper. It is worth to note that in
the sum (\ref{qf}) $V_{\bm k - \bm p}^2 +  V_{\bm k - {\bm p}'}^2$
can be replaced by $2 V_{\bm k - \bm p}^2$.

{\it Electrons polarized along the same axis}. By using the
coordinate system with $z$ parallel to the electron spin
polarization one has $s_{{\bm k}, x} = s_{{\bm k}, y} \equiv 0$
and the spin-density matrix is a diagonal matrix with the diagonal
components $f_{{\bm k}, s}$ ($s = \pm 1/2$). It follows then that
the products $\hat \rho^{(1)}(\bm{k}) \hat \rho^{(2)}(\bm{k}')$
and $[ I^{(1)} - \hat \rho^{(1)}(\bm{p}) ] [ I^{(2)} - \hat
\rho^{(2)}(\bm{p}') ]$ are diagonal as well. We can take into
account the spin conservation
\[
M({\bm p}, s_3; {\bm p}', s_4| {\bm k}, s_1; {\bm k}', s_2)
\propto \delta_{s_3 + s_4, s_1 + s_2}
\]
and use the identity
\[
M({\bm p}, s_3; {\bm p}', s_4| {\bm k}, s_1; {\bm k}', s_2) M({\bm
p}, s_3; {\bm p}', s_4| {\bm k}, - s_1; {\bm k}', s_2) = 0\:.
\]
This allows, in agreement with Ref.~\onlinecite{amico}, rewrite
the collision term for $f_{{\bm k}, s}$ as
\[
\frac{2 \pi}{\hbar} \sum_{{\bm k}' {\bm p} {\bm p}'} \sum_{s' s_1
s_2 }\delta_{{\bm k} + {\bm k}',\: {\bm p} + {\bm p}'}\:
\delta(E_k + E_{k'} - E_p - E_{p'}) W({\bm p}, s_1; {\bm p}', s_2|
{\bm k}, s; {\bm k}', s')
\]
\[
\times [f_{{\bm k}, s}f_{{\bm k}', s'}(1-f_{{\bm p}, s_1})
(1-f_{{\bm p}', s_2 }) - f_{{\bm p}, s_1} f_{{\bm p}', s_2 }
(1-f_{ {\bm k}, s} ) (1-f_{{\bm k}', s'})]\:.
\]
Here
\begin{eqnarray}
&&W({\bm p}, s; {\bm p}', s| {\bm k}, s; {\bm k}', s) = (A + B)^2
= (V_{\bm k - \bm p} - V_{\bm k - {\bm p}'})^2\:, \nonumber \\
&&W({\bm p}, s; {\bm p}', - s| {\bm k}, s; {\bm k}', - s) = (A
-B)^2 = V_{\bm k - \bm p}^2 \:,\nonumber \\ &&W({\bm p}, - s; {\bm
p}', s| {\bm k}, s; {\bm k}', - s) = (2 B)^2 = V_{\bm k - {\bm
p}'}^2 \nonumber \:,
\end{eqnarray}
and other values of $W$ with $s_1+s_2 \ne s+s'$ are zero.

{\it Low electron polarization}. If the average electron spin
${\bm s}_{\bm k}$ is small as compared with the occupation
probability $f_{\bm k}$ then, in the equation for $f_{\bm k}$, one
can ignore the spin polarization at all and use Eq.~(\ref{qf}),
while, in the equation for ${\bm s}_{\bm k}$, one can retain in
the collision term only the contribution linear in ${\bm s}_{\bm
k}$. The linearized collision term is given by
\begin{equation} \label{q_fd}
\bm{Q}_{\bf k} \{\bm s, f\} = \frac{2\pi}{\hbar} \sum_{\bm k', \bm
p, \bm p'} \delta_{\bm k+ \bm k', \bm p+ \bm p'}\: \delta (E_k +
E_{k'} - E_p - E_{p'})
\end{equation}
\[
\times \left[ (V_{\bm k - \bm p}^2 + V_{\bm k - {\bm p}'}^2 -
V_{\bm k - \bm p} V_{\bm k - {\bm p}'}) {\bm s}_{\bm k} F({\bm
k}';{\bm p},{\bm p}') - V_{\bm k - \bm p} V_{ \bm k - {\bm p}'}
{\bm s}_{{\bm k}'} F({\bm k};{\bm p},{\bm p}')  \right.
\]
\[
\left. - V_{\bm k - \bm p} (V_{\bm k - \bm p} - V_{\bm k - {\bm
p}'}) {\bm s}_{\bm p} F({\bm p}';{\bm k},{\bm k}') - V_{\bm k -
{\bm p}'} (V_{\bm k - {\bm p}'} - V_{\bm k - \bm p}) {\bm s}_{{\bm
p}'} F({\bm p};{\bm k},{\bm k}')  \right] \:,
\]
where
\[
F({\bm k}_1;{\bm k}_2,{\bm k}_3) = f_{{\bm k}_1} (1 - f_{{\bm
k}_2})(1 - f_{{\bm k}_3}) + (1 - f_{{\bm k}_1})  f_{{\bm k}_2}
f_{{\bm k}_3} = f_{{\bm k}_1} (1 - f_{{\bm k}_2} - f_{{\bm k}_3})
+  f_{{\bm k}_2} f_{{\bm k}_3}\:.
\]
Equation (\ref{q_fd}) can be transformed to
\begin{equation}\label{q_fd1}
\bm{Q}_{\bf k} \{\bm s, f\} = \frac{2\pi}{\hbar} \sum_{\bm k', \bm
p, \bm p'} \delta_{\bm k+ \bm k', \bm p+ \bm p'}\: \delta (E_k +
E_{k'} - E_p - E_{p'})\hspace{3 cm}\mbox{}
\end{equation}
\[
\times \left\{ 2 V_{\bm k - \bm p}^2 \left[ {\bm s}_{\bm k} F({\bm
k}';{\bm p},{\bm p}') - {\bm s}_{\bm p} F({\bm p}';{\bm k},{\bm
k}') \right] \right. - \hspace{2 cm}\mbox{}
\]
\[
\mbox{}\hspace{1.5 cm} \left.- V_{\bm k - \bm p}V_{\bm k - \bm p'}
\left[ {\bm s}_{\bm k} F({\bm k}';{\bm p},{\bm p}')  + {\bm
s}_{\bm k'} F({\bm k}';{\bm p},{\bm p}') - 2 {\bm s}_{\bm p}
F({\bm p}';{\bm k},{\bm k}') \right] \right\}.
\]
Here the term proportional to $2 V_{\bm k - \bm p}^2$ is due to
the direct Coulomb interaction whereas the term proportional to
$V_{\bm k - \bm p}V_{\bm k - \bm p'}$ comes from the exchange
interaction.

{\it Non-degenerate 2D electron gas}.  In this case the function
$F({\bm k}_1;{\bm k}_2,{\bm k}_3)$ reduces to $f_{{\bm k}_1}$ and
the collision terms are as follows
\begin{equation}
Q_{\bm k} \{ f, \bm{s} \} = \frac{2 \pi} {\hbar}\sum_{{\bm k}'
{\bm p} {\bm p}'} \delta_{{\bm{\scriptstyle k+k'}},{{\bm p+ \bm
p'}}}\:\delta(E_k+E_{k'}-E_p-E_{p'})
\end{equation}
\[
\times [( 2 V^2_{{\bm k}-{\bm p}} - V_{{\bm k}-{\bm p}} V_{{\bm
k}-{\bm p}'}) ( f_{\bm k} f_{{\bm k}'} - f_{\bm p} f_{{\bm p}'}) -
V_{{\bm k}-{\bm p}} V_{{\bm k}-{\bm p}'} ({\bm s}_{\bm k} \cdot
{\bm s}_{{\bm k}'} - {\bm s}_{\bm p} \cdot {\bm s}_{{\bm p}'})]\:,
\]
\begin{equation} \label{exchange}
\bm{Q}_{{\bm k}} \{ \bm{s}, f \} = \frac{2\pi}{\hbar}\sum_{{\bm
k}' {\bm p} {\bm p}'} \delta_{{\bm{\scriptstyle k+k'}},{{\bm p+
\bm p'}}}\:\delta(E_k+E_{k'}-E_p-E_{p'})
\end{equation}
\[
\times [2V_{{\bm k}-{\bm p}}^2( {\bf s}_{\bm k} f_{{\bm k}'} -
{\bf s}_{\bm p} f_{{\bm p}'})- V_{{\bm k}-{\bm p}} V_{{\bm k}-{\bm
p}'}( {\bf s}_{\bm k} f_{k'} +{\bf s}_{{\bm k}'} f_{{\bm k}} -
2{\bf s}_{\bm p} f_{{\bm p}'})]\:.
\]
Neglecting the exchange interaction given by the term proportional
to the $V_{{\bm k}-{\bf p}}V_{{\bm k}-{\bf p}'}$ the scattering
rate $Q_{\bm k} \{ f, \bm{s} \}$ is independent on the electron
spin distribution. Both $Q_{\bm k} \{ f, \bm{s} \}$ and
$\bm{Q}_{{\bm k}} \{ \bm{s}, f \}$ take a simple form
\begin{equation} \label{eeintegral}
Q_{{\bm k}} \{ f, \bm{s} \} = \sum_{{\bm k}' {\bm p} {\bm p}'}
W_{{\bm p \bm p}',{\bm k \bm k}'} ( f_{\bm k} f_{{\bm k}'} -
f_{\bm p} f_{{\bm p}'})\:,\:\bm{Q}_{{\bm k}} \{ \bm{s}, f^0 \} =
\sum_{{\bm k}' {\bm p} {\bm p}'} W_{{\bm p \bm p}',{\bm k \bm k}'}
( {\bf s}_{\bm k} f_{{\bm k}'} - {\bm s}_{\bm p} f_{{\bm p}'})\:,
\end{equation}
where $W_{{\bm p \bm p}',{\bm k \bm k}'}$ is the probability rate
for the scattering of a pair of electrons from the ${\bm k}, {\bm
k}'$ states to the ${\bm p}, {\bm p}'$ states
\[
W_{{\bm p \bm p}',{\bm k \bm k}'} =\frac{2 \pi} {\hbar}
\delta_{{\bm{\scriptstyle k+k'}},{{\bm p+ \bm
p'}}}\:\delta(E_k+E_{k'}-E_p-E_{p'}) \:2\:V_{{\bm k}-{\bf p}}^2\:,
\]
an additional factor of 2 takes into account the double degeneracy
of the electronic states.
\section{Electron-Electron Scattering Time Controlling the DPSR}
In what follows we consider only a non-degenerate 2D electron gas
in which case the zero-approximation spin-density matrix
(\ref{rho}) can be written as
\begin{equation}
\hat \rho^0_{\bm k}= f^0_{\bm k} (1 + 2 \bar{{\bm s}} \cdot {\bm
\sigma})\:,
\end{equation}
where $f^0_{\bm k}$ is the Boltzmann distribution function, and
$\bar{{\bm s}}$ is the average spin per electron, ${\bm S}/(n_+ +
n_-)$. The nonequilibrium correction $\delta {\bm s}_{\bm k }$
satisfies the equation
\begin{equation}\label{eq_corr}{\bm
\Omega}_{\bm k} \times (2 f^0_{{\bm k}} \bar{{\bm s}}_0) + Q_{\bm
k} \{ \delta {\bm s}, f^0 \} = 0\:.
\end{equation}
It should be noted that in the 2D case the collision term
(\ref{exchange}) does not allow the quasi-elastic and relaxation
time approximations and Eq. (\ref{eq_corr}) must be solved
directly.

The $\alpha$-component of the vector product ${\bm \Omega}_{\bm k}
\times \bar{{\bm s}}_0$ can be represented as $({\bm \Omega}_{\bm
k} \times \bar{{\bm s}}_0)_{\alpha}=\Lambda_{\alpha \beta \gamma}
k_{\beta} \bar{s}_{0\gamma},$ where the third-rank tensor ${\bm
\Lambda}$ in general case of both BIA and SIA linear-${\bm k}$
terms has four nonzero components
\[
\Lambda_{xxz} = -\Lambda_{zxx} = \beta_+/\hbar, \quad
\Lambda_{yyz} = -\Lambda_{zyy} = \beta_-/\hbar\:.
\]

The function $(1/k_{\beta}) Q_{\bm k}\{ k_{\beta} F_k , f^0 \}$ is
independent of the azimuthal angle $\Phi_{\bm k}$ (here $F_k$ is
an arbitrary function of $k = |{\bm k}|$) as the operator
$Q_{\bm k} \{ \delta s_{\alpha}, f^0 \}$ conserves the angular
distribution in the ${\bm k}$ space. In such case the solution may
be written as follows
\begin{equation} \label{vk}
\delta s_\alpha ({\bm k}) = - \Lambda_{\alpha \beta \gamma}
\frac{k_\beta}{k} \bar{s}_{0\gamma}\: k_T\:
\tau^*_{ee}\:e^{\mu/k_B T}\: v(K)\:.
\end{equation}
Here we introduced the dimensionless wavevector ${\bf K} = {\bm
k}/k_T\:, \: k_T = (2mk_B T/\hbar^2)^{1/2}$,
\begin{equation}
\tau^*_{ee} = \frac{\hbar k_B T \ae^2}{e^4N}\:,
\end{equation}
$N=n_+ + n_-$ and $v(K)$ satisfies the equation
\begin{equation} \label{nounit}
K e^{-K^2}=\int d^2K'\int d^2P \; \widetilde W_{{\bf PP}',{\bf
KK}'} \left(v(K)e^{-K'^2}-\cos{\Theta} \; v(P)e^{-P'^2}\right)\:,
\end{equation}
where $\Theta$ is the angle between ${\bf K}$ and ${\bf P}$, $\bf
P' = K + K' - P$,
\[
\widetilde W_{{\bf PP}',{\bf KK}'} = \frac{1}{|\mathbf K - \mathbf P|^2}
\:\:\delta(K^2+K'^2-P^2-P'^2)\:.
\]

Inserting Eq.~(\ref{vk}) into Eq.~(\ref{balances0}) one obtains
after summation over ${\bm k}$ the principal values of the tensor
of reciprocal spin relaxation times
\begin{equation} \label{eetime}
\frac{1}{\tau^s_{x'x'}} = \left( \frac{\beta_+ k_T}{\hbar}
\right)^2 \tau\:, \qquad \frac{1}{\tau^s_{y'y'}} = \left(
\frac{\beta_- k_T}{\hbar} \right)^2 \tau\:, \qquad
\frac{1}{\tau^s_{zz}} = \frac{1}{\tau^s_{x'x'}} +
\frac{1}{\tau^s_{y'y'}}\:.
\end{equation}
The time $\tau$ which controls the spin relaxation is given by
\begin{equation} \label{itau}
\tau= \tau^*_{ee} I, \qquad I = \frac{1}{2}\int\limits_0^\infty
v(K) K^2 dK \:.
\end{equation}
The parameter $\tau^*_{ee}$ is present also in the $ee$-scattering
time which determines the rate of energy exchange between $2D$
electrons Ref.~\onlinecite{esipov}.

The function $v(K)$ was expanded in series using a basis set
$l_n({\varepsilon}) =$ $\sqrt{2} \exp{(- \varepsilon)}
L_n(2\varepsilon)$, where $L_n(\varepsilon)$ are the Laguerre
polynomials and $\varepsilon = K^2$. The expansion was substituted
into the right-hand side of Eq. (\ref{nounit}), the integration
has been performed by Monte-Carlo method. The problem was reduced
to a set of linear inhomogeneous equations for the expansion
coefficients of $v(K)$. The resulting value of $I$ in
Eq.~(\ref{itau}) was found to be $\approx 0.027$. Allowance for
the exchange interaction leads to a slight increase of this value
to $\approx 0.028$.

{\it Comparison with ionized impurities scattering}. If the
ionized impurities of the same concentration $N$ lie inside the 2D
layer then the corresponding transport time is given by $\tau_{tr}
= (2/\pi^2) \tau^*_{ee}$ (see Ref.~\onlinecite{lyo}). The spin
relaxation time, $\tau^{ii}_s$, controlled by scattering by the
ionized impurities is given by Eq.~(\ref{eetime}) where $\tau$ is
changed by $\tau_{tr}/2$, thus the ratio of the spin relaxation
time, $\tau^{ee}_s$, governed by electron-electron collisions and
the time $\tau^{ii}_s$ is $\approx 3.6$, i.e. the elastic
scattering by impurities is less efficient. If the doped layer is
separated from the quantum well by a spacer, the influence of the
Coulomb potential of ionized impurities on $\tau_s$ is reduced and
eventually can be neglected.

The above result was obtained for the 2D Coulomb potential $
U^{2D}(\rho) = e^2 / \ae \rho$, where $\rho$ is the distance
between the electrons in the interface plane. It is this potential
that leads to the Fourier transform given by Eq.~(\ref{2d}). In
order to analyze the role of the quasi-2D character of the
electron wave function confined in a QW of the finite thickness
$a$ one can replace $U^{2D}(\rho)$ by the effective potential
obtained by averaging the three-dimensional Coulomb potential as
follows
\[
U(\rho) = \frac{e^2}{\ae} \int\int \frac{\varphi^2_{e1}(z)
\varphi^2_{e1}(z')}{\sqrt{\rho^2 + (z-z')^2}}\: dz dz'\:,
\]
where $\varphi_{e1}(z)$ is the electron envelope function at the
lowest conduction subband $e1$. The straightforward calculation
shows Ref.~\onlinecite{glazov} that for the conduction band offset
$V_c\gg\hbar^2/ma^2$ and $k_Ta<1$ a value of the time $\tau =
\tau^*_{ee} I$ increases with widening the QW as the
electron-electron interaction becomes weaker, but an order of
magnitude of $I$ and $\tau$ remains the same as in the exact 2D
case.

\section*{Conclusion. Future Work}
We have shown that electron-electron collisions control the
D'yakonov-Perel' spin relaxation in the same way as any scattering
processes do. The electron-electron collision integral has been
derived for an arbitrary degeneracy and spin distribution of the
2D electron gas. The calculations has been performed for
non-degenerate 2D and quasi-2D electrons confined in a quantum
well. Calculations of the spin relaxation time controlled by
electron-electron collisions in bulk semiconductors are in
progress. An important next step is the extension of the
calculations from non-degenerate to degenerate spin-polarized
electron gas.

In agreement with our theory, the latest optical spin-dynamic
measurements in an $n$-doped\\ GaAs/AlGaAs QW of high mobility
give an experimental evidence of the electron-electron scattering
to randomize the electron spin precession
Ref.~\onlinecite{harley1}.

Note in conclusion that the time $\tau = \tau^*_{ee} I$ can be
related physically with the momentum relaxation time of an
electron by equilibrium holes of the density $N$ if the electron
and hole effective masses are assumed to coincide. In the electron
collisions with holes the directed electron momentum is
transferred to the hole gas and decays within the time $\sim
\tau$.  The work was supported by RFBR, and by the programs of the
Russian Ministry of Sci. and the Presidium of the Russian Academy
of Sci.

\end{document}